\documentclass[graybox]{svmult}

\usepackage{mathptmx}       % selects Times Roman as basic font
\usepackage{helvet}         % selects Helvetica as sans-serif font
\usepackage{courier}        % selects Courier as typewriter font
\usepackage{type1cm}        % activate if the above 3 fonts are
\usepackage{makeidx}         % allows index generation
\usepackage{graphicx}        % standard LaTeX graphics tool
\usepackage{multicol}        % used for the two-column index
\usepackage[bottom]{footmisc}% places footnotes at page bottom

\newcommand{\cmt}[1]{}

\def\cm3{~{\rm cm^{-3}}}
\newcommand{\rxj}{RX J1713.7-3946}
\def\lesssim{\buildrel < \over {_{\sim}}}

\makeindex             % used for the subject index
\begin{document}

\title*{New insights on hadron acceleration at supernova remnant shocks}
% Use \titlerunning{Short Title} for an abbreviated version of
% your contribution title if the original one is too long
\author{Damiano Caprioli}
% Use \authorrunning{Short Title} for an abbreviated version of
% your contribution title if the original one is too long
\institute{Damiano Caprioli \at Princeton University, 4 Ivy Ln, 08544 Princeton (NJ), US. \email{caprioli@astro.princeton.edu}}
%
% Use the package "url.sty" to avoid
% problems with special characters
% used in your e-mail or web address
%
\maketitle

\abstract*{We outline the main features of nuclei acceleration at supernova remnant forward shocks, stressing the crucial role played by self-amplified magnetic fields in determining the energy spectrum observed in this class of sources.
In particular, we show how the standard predictions of the non-linear theory of diffusive shock acceleration has to be completed with an additional ingredient, which we propose to be the enhanced velocity of the magnetic irregularities particles scatter against, to reconcile the theory of efficient particle acceleration with recent observations of gamma-ray bright supernova remnants.}

\abstract{We outline the main features of nuclei acceleration at supernova remnant forward shocks, stressing the crucial role played by self-amplified magnetic fields in determining the energy spectrum observed in this class of sources.
In particular, we show how the standard predictions of the non-linear theory of diffusive shock acceleration has to be completed with an additional ingredient, which we propose to be the enhanced velocity of the magnetic irregularities particles scatter against, to reconcile the theory of efficient particle acceleration with recent observations of gamma-ray bright supernova remnants.}

\section{Introduction}
Supernova remnants (SNRs) have been regarded for many decades now as the sources of Galactic cosmic rays (CRs), both because of their energetics and of the fact that strong shocks are expected to naturally accelerate particles with power-law energy distributions, according to the so-called Fermi mechanism.

Clear-cut evidences of particle acceleration in SNRs have been found many years ago in radio to X-ray synchrotron emission, therefore attesting the presence of \emph{electrons} with energies as high as 10-100 TeV. 
On the other hand, direct evidences of \emph{proton} acceleration are much less clear: the most prominent signature of hadronic acceleration has been individuated in the $\gamma$-ray emission from the decay of neutral pions produced in nuclear interaction between relativistic nuclei and the interstellar medium \cite{dav94}.
The first SNRs detected in the TeV $\gamma$-rays, however, did not unravel the question, since in this energy range  it is hard to disentangle whether the observed spectrum (typically exhibiting a cut-off around $\sim$10 TeV) is the result either of the hadronic mechanism described above or of inverse-Compton scattering (ICS) of relativistic electrons on some photon background.

Coupling the TeV observations of Cherenkov telescopes as HESS, VERITAS and MAGIC with the GeV observations of the Fermi satellite, however, it has been possible to assess the nature of the emission, at least in some paradigmatic cases: SNR \rxj~ turned out to be mostly consistent with a leptonic scenario \cite{Fermi1713} while, on the contrary, Tycho's SNR has been convincingly modeled according to the hadronic hypothesis \cite{tycho}.
The main element allowing such a distinction is \emph{the slope of the observed $\gamma$-ray spectrum}: for a $\propto E^{-q}$ CR energy distribution, ICS is expected to produce a flatter power-law photon spectrum, 
$\propto E_{\gamma}^{-(q+1)/2}$, 
while pion decay and bremsstrahlung produce a spectrum parallel to the parents' one, $\propto E_{\gamma}^{-q}$. 

\begin{figure}[b]
\sidecaption[t]
\includegraphics[scale=.6]{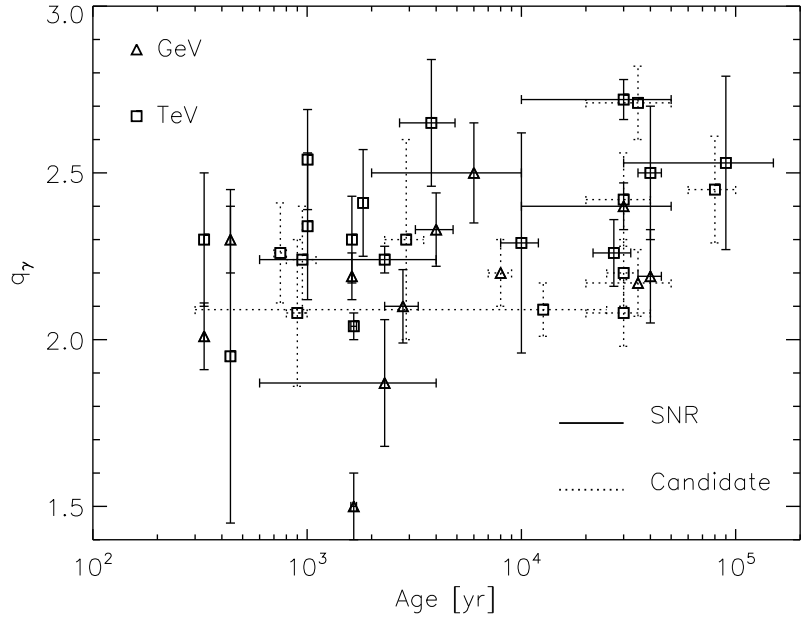}
\caption{Slopes $q_\gamma$ of the $\gamma$-ray spectra as inferred in the GeV/TeV band for both confirmed and candidate SNRs, as in the legend. Data from table 1 in ref.~\cite{gamma}.}
\label{fig:slopes}       
\end{figure}

Both GeV and TeV observations are ---in basically all the cases--- consistent with spectra steeper than $E^{-2}$, as showed in Fig.~\ref{fig:slopes} (see ref.~\cite{gamma} for more details).
The only SNRs showing a GeV slope smaller than 2 are in fact the aforementioned \rxj~ and Vela Jr., whose emission may be accounted for by invoking a strong photon background making ICS dominate over pion decay.
When the spectrum is steeper than $E^{-2}$ across many energy decades, however, a hadronic mechanism is favored, instead, since bremsstrahlung is typically negligible in SNRs.

But this behavior has another important consequence in terms of the process responsible for particle acceleration (of both electrons and protons, which share the same spectrum in rigidity): it has to be able to produce particle spectra steeper than $E^{-2}$. 
This apparently harmless requirement is at odds with any theory of \emph{efficient} particle acceleration at SNRs developed in the last 30 years since, according to Fermi's mechanism, the expected slope is a function of the shock compression ratio only, and namely $q=\frac{r+2}{r-1}$. 
For strong shocks $r=\frac{\gamma+1}{\gamma-1}=4$, with $\gamma=5/3$ the gas adiabatic index and, in turn, $q=2$. 
When acceleration is efficient, though, one could expect the shock to be modified by the back-reaction of the accelerated particles, whose pressure may become comparable with the ram one. 
The simplest way to account for this effect is considering that CRs are relativistic particles forming a gas with adiabatic index $\gamma_{cr}=4/3$, which would provide a compression ratio $r=7$ and in turn $q=1.5$. 
The escape of high-energy particles may in addition make the shock behave as partially radiative, increasing $r$ even further and pushing $q$ down to $1-1.2$.
In any case, a very solid prediction of the \emph{non-linear theory of diffusive shock acceleration} (NLDSA) is that the CR spectrum has to be invariably flatter than $E^{-2}$ at the highest energies, and being flatter and flatter for larger and larger acceleration efficiencies (for more details on CR modified shocks see, e.g., ref.~\cite{malkov-drury01}).

Therefore, independently of the possible origin (leptonic or hadronic) of the observed GeV emission, Fermi-LAT's observations force us to rethink our theory of CR acceleration, challenging the paradigm which requires SNR shocks to channel fraction as large as 10-30\% of their kinetic energy into accelerated particles in order to be the potential sources of Galactic CRs.

\section{Magnetic field amplification and particle acceleration}  
A fundamental tile in the mosaic of the comprehension of particle acceleration in (young) SNRs has also come from the detection of narrow X-ray bright rims immediately downstream of their forward shocks. 
The non-thermal nature of their spectra has been accounted for as due to synchrotron emission of relativistic electrons, while their narrowness (typically less than 0.01pc) points to magnetic fields as large as a few hundreds $\mu$G, almost two orders of magnitude larger than the typical interstellar one.
Such an evidence has been welcomed by theorists for several reasons.
\begin{itemize}
\item The super-Alfv\`enic streaming of CRs upstream of the shock has been predicted to lead to the excitation of several magnetic modes via plasma instabilities \cite{skilling75a, bell04}. 
\item A higher level of magnetization enhances particle diffusion and allows the achievement of larger energies, arguably up to a few PeV, namely enough to account for the knee observed in the Galactic CRs detected at Earth \cite{bac07}.
\item The pressure in magnetic turbulence may be so large to overcome the gas one, therefore preventing the CR pressure to modify the shock too severely \cite{jumpl}.
\item Finally, magnetic field self-generation may have a key role also in determining the CR spectral slope, and therefore in reconciling NLDSA theory with $\gamma$-ray observations, which is what we want to outline here (also see ref.~\cite{efficiency}).
\end{itemize}
Without delving into the details of a kinetic theory of plasmas, we can state that, being the accelerated particles and the magnetic fields coupled through resonant scattering, the whole system has to gradually reduce the relative velocity of CRs, whose diffusion velocity is basically the shock velocity in the upstream reference frame, $u$, and of magnetic irregularities, whose phase velocity is of the order of the Alfv\`en speed, $v_A=B/\sqrt{4\pi\rho}$.
In other words, magnetic field amplification might be considered as an inevitable consequence of an efficient CR acceleration while, on the other hand, an enhanced level of magnetization is a necessary condition for an efficient particle diffusion and acceleration: if the initial Alfv\`enic Mach number $M_A=u/v_A$ is large (as it typically is is in the interstellar magnetic field), it tends to be reduced by a non-linear interplay between particles and fields.

\begin{figure}[b]
\includegraphics[scale=.65]{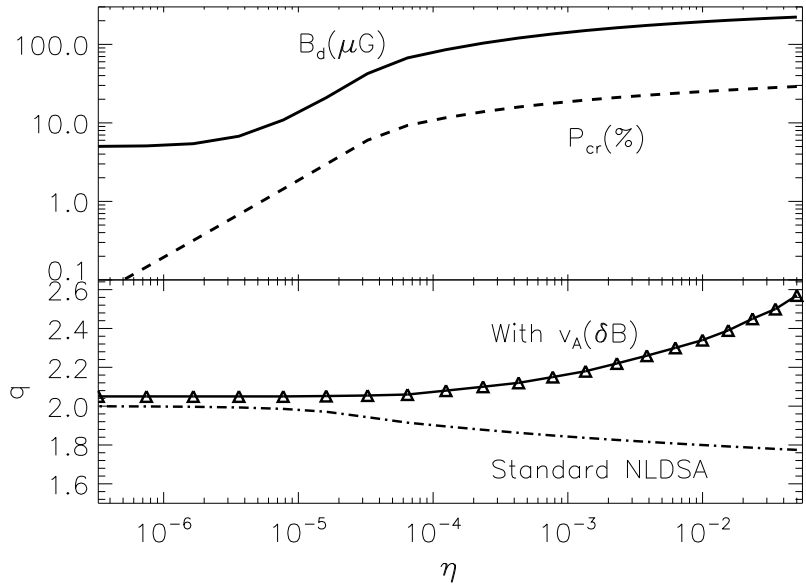}
\caption{\emph{Top panel}: pressure in CRs at the shock, $P_{cr}$, expressed as a fraction of the ram pressure $\rho u^2$, and predicted downstream magnetic field, $B_d$, as a function of the injection efficiency $\eta$.
The environmental parameters are those relevant for a typical SNR in the early Sedov stage (see ref.~\cite{efficiency} for more details).
\emph{Bottom panel}: CR spectral slope obtained when the Alfv\`en velocity is calculated in the self-amplified magnetic field $\delta B$ (solid line with triangles) or according to a standard NLDSA theory assuming $M_A(B_0)\gg 1$ (dot-dashed line).}
\label{fig:eff}      
\end{figure}

We model such an interplay by adopting the semi-analytic formalism for NDLSA outlined in ref.~\cite{efficiency} and references therein, which proved to be quick but very accurate for non-relativistic shocks \cite{comparison}.
We solve the equations of hydrodynamics coupled with a kinetic description of the non-thermal particles (via the diffusion-convection equation, eq.\ 2.5 in ref.~\cite{efficiency}) and a fluid description for the generation of magnetic turbulence via resonant streaming instability (eq.\ 2.18 in ref.~\cite{efficiency}).
The dynamical feedback of both accelerated particles and amplified magnetic fields are self-consistently retained in this approach, as well as the escape of the highest energy particles from the upstream boundary \cite{escape}.
\begin{svgraybox}

The key role of magnetic field amplification in determining the CR spectral slope is to enhance the velocity of the scattering centers, i.e., the phase velocity of the magnetic waves, $\sim v_A\propto \delta B$. 
Since the compression ratio shaping the CR spectra is not the one of the fluid, but rather the one of the velocity of their scattering centers ahead and behind the shock, when $v_A$ becomes a non-negligible fraction of $u$, it is in principle possible for diffusive shock acceleration to produce spectra steeper than $E^{-2}$.\end{svgraybox}

More precisely, the compression ratio felt by the fluid between upstream and downstream is the usual $r=u_u/u_d\geq 4$ for strong, CR modified shocks, while the compression ratio felt by CRs is instead $\tilde{r}=(u_u+v_{A,u})/(u_d+v_{A,d})$.
According to the argument above, however, the coupling between CRs and magnetic field is expected to produce waves streaming \emph{against} the fluid in the upstream (in the direction opposite to the CR gradient), while downstream the turbulence should be efficiently isotropised ($v_{A,d}\sim 0$). 

The net effect is that $\tilde{r}=(1-M_{A,u}^{-1})r$ and, in turn, if $M_A(\delta B)$ becomes small enough, the predicted slope of the CR spectrum $q=\frac{\tilde{r}+2}{\tilde{r}-1}$ may become larger than 2, even for strong shocks and large CR pressures. 
On the other hand, a steeper CR spectrum allocates less pressure (which in turn generates less magnetic field), therefore we need a non-linear calculation to understand which effect dominates the determination of the spectral slope when acceleration is efficient: the pressure in CRs, which makes the shock more compressible, or rather the reduced jump in $\tilde{r}$.

Since CR injection has not been understood from first principles, yet, we cannot but parametrize it: we allow $\eta$, i.e., the fraction of particle crossing the shock with a momentum large enough to be injected into the acceleration mechanism, to vary between $\sim 10^{-7}$ and $\sim 0.1$.
Such a large range of injection efficiency reflects different regimes of CR acceleration, spanning from unmodified shocks (where the pressure in CRs, $P_{cr}$ is expected to be negligible with respect to the ram pressure), to strongly modified ones, in which the shock structure is deeply affected by CRs and in which magnetic field amplification is very effective.
At saturation, in fact, we have that $P_{B}=\delta B^2/8\pi\sim P_{cr}/M_A$ (see \S2.1 in ref.~\cite{efficiency} for the exact dependence).

Our results are showed in Fig.~\ref{fig:eff} for a SNR about 3000 yr old, i.e., in the early Sedov stage.
For a much wider discussion of the SNR evolution in terms of its non-thermal content the reader may refer to refs.~\cite{crspectrum,efficiency}.
In the top panel, $P_{cr}$ and the downstream field $B_d$ are showed as function of $\eta$: they reproduce the test-particle case for $\eta\lesssim 10^{-5}$, show a linear increase for about a decade and finally saturate to about 10-20\% and 100-200$\mu$G, respectively, above $\eta\sim 10^{-4}$.

The very reason of this saturation, which among other things assesses the mild dependence of our findings on $\eta$, is that the whole system reacts to high injection efficiencies by steepening the CR spectrum in order to accommodate almost the same energy in non-thermal particles, as demonstrated in the bottom panel of Fig.~\ref{fig:eff} (solid line).
Spectra as steep as $E^{-2.2}-E^{-2.5}$ correspond to the highest acceleration efficiencies and to the fields inferred in the downstream of young shell SNRs (see figure 3 in ref.~\cite{efficiency}).
The spectral slope obtained is also compared with the one predicted by a two-fluid theory of NLDSA, $q=(3\gamma_{eff}-1)/2$, where the CR contribution is accounted for via a compression ratio modulated by an effective adiabatic index of the gas+CR fluid \cite{chevalier83}, namely $\gamma_{eff}=\frac{5+3P_{cr}}{3-3P_{cr}}$ (dot-dashed line in the bottom panel of Fig.~\ref{fig:eff}). 
It is worth recalling that the latter slope, already clearly inconsistent with the observations reported in Fig.~\ref{fig:slopes}, represents an \emph{upper limit} to the one which would be predicted by taking into account also the escape of CRs, which would make the shock partially radiative, in principle increasing $r$ well above 7 \cite{malkov-drury01}.

\begin{svgraybox}
The only way to accommodate the typical efficiencies (10-20\%) required for SNRs to be the sources of Galactic CRs, and the high levels of magnetization observed (a few 100$\mu$G), with CR spectra steeper than $E^{-2}$ is to modify the standard theory of NLDSA.

We propose here that accounting for the features of the magnetic field produced by the CR streaming is a possible way to build a consistent picture of efficient NLDSA in SNRs, actually reversing the usual trend which relates larger and larger CR acceleration efficiencies with flatter and flatter spectra.   
\end{svgraybox}
%The effect we discussed, while rather natural in terms of CR--magnetic field interplay, needs indeed to be checked against first-principle simulations of collisionless shocks with environmental parameters relevant for SNRs.
%Preliminary results, nevertheless, show that magnetic field amplification at parallel shocks usually produces an increase in the effective Alf\'en velocity \cite{rs10,gs12}.

\bibliographystyle{spmpsci}
\bibliography{bibeff}
\end{document}